\documentclass[12pt]{article}
\usepackage{times}
\usepackage{graphicx} 
\usepackage{float} 
\usepackage{bm} 
\usepackage{cite}
\usepackage{xcolor}
\usepackage{physics}

\title{Processing entangled photons in high dimensions with a programmable light converter}
\author
    {Ohad Lib, Kfir Sulimany and Yaron Bromberg$^\ast$\\
\\
\normalsize{Racah Institute of Physics, The Hebrew University of Jerusalem, Jerusalem, 91904 Israel}\\
\\
\normalsize{$^\ast$To whom correspondence should be addressed; E-mail:  Yaron.Bromberg@mail.huji.ac.il.}
}

\date{}

\begin{document} 

\maketitle 

\begin{abstract}
High-dimensional entanglement offers a variety of advantages for both fundamental and applied applications in quantum information science. A central building block for such applications is a programmable processor of entangled states, which is crucial for the certification, manipulation and distribution of high-dimensional entanglement. The leading technology for processing photons is integrated multiport interferometers. However, such devices are incompatible with structured light, and their scaling is challenging. Here, we unlock these limitations by demonstrating a reconfigurable processor of entangled photons in high-dimensions that is based on multi-plane light conversion (MPLC), a technology that was recently developed for multiplexing hundreds of spatial modes for classical communication. We use our programmable MPLC platform to certify three-dimensional entanglement in two mutually unbiased bases, perform 400 arbitrary random transformations on entangled photons, and convert the mode basis of entangled photons for entanglement distribution.
\end{abstract}

\maketitle 

\section{Introduction}

High-dimensional entanglement is attracting increasing attention in quantum information science\cite{friis2019entanglement,erhard2020advances}. States entangled in high dimensions can exhibit stronger correlations\cite{vaziri2002experimental,thew2004bell}, increase the information density in quantum communication\cite{ali2007large}, provide superior computational resources\cite{reimer2019high}, and are more resilient to noise and loss\cite{ecker2019overcoming}. Taking advantage of such entangled states for both fundamental and applied applications requires efficient methods to generate, manipulate, certify, and distribute high-dimensional entanglement\cite{friis2019entanglement,erhard2020advances}.

While the generation of high-dimensional entangled states has been demonstrated over a variety of photonic platforms and degrees of freedom\cite{avenhaus2009experimental,dada2011experimental,bavaresco2018measurements,valencia2020high,kysela2020path}, platforms for scalable and universal processing of entangled photons, which are required for the certification, manipulation and distribution of entanglement, are lacking. To date, the leading platform for universal processing of entangled states is photonic integrated circuits, where the input photons encoded in $N$ waveguide modes are transformed via a mesh of $\sim N^2$ integrated Mach-Zehnder interferometers with programmable phase shifters\cite{o2007optical,o2009photonic,matthews2009manipulation,carolan2015universal,flamini2018photonic,wang2018multidimensional,taballione2021universal,larocque2021englund}. The quadratic scaling with the number of modes makes the integrated approach sensitive to fabrication errors and interferometric noise. Moreover, integrated processors are incompatible with encoding high-dimensional bits of information in the transverse modes of photons, thus precluding fundamental tests and applications of entanglement of structured photons \cite{van2014ultra,mirhosseini2015high,sit2017high,erhard2020advances,piccardo2021roadmap}. Recently, a different approach compatible with high-dimensional encoding with transverse modes has been proposed and demonstrated, using wavefront shaping and a multi-mode fiber as a reconfigurable multimode processor\cite{leedumrongwatthanakun2020programmable,goel2022inverse}. While being a promising direction for studying linear quantum networks, it requires careful characterization of the complex transmission matrix of the multimode fiber, which limits the available optical bandwidth and temporal stability\cite{leedumrongwatthanakun2020programmable,matthes2019optical}.

Here, we consider multi-plane light conversion (MPLC), a technology developed for multimode processing in classical optical communication, as a universal platform for processing high dimensional entanglement. MPLC is based on performing general and reconfigurable unitary transformations on arbitrary spatial modes, using only a few optimized phase masks separated by free-space propagation\cite{morizur2010programmable,labroille2014efficient,song2021simultaneous}. MPLC has several unique features that make it attractive for processing entangled photons. It exhibits high phase stability due to its common path configuration, it supports arbitrary input and output modes, and it offers extremely wide-band operation\cite{fontaine2020ultrabroadband,hiekkamaki2021high}. Most importantly, the number of phase masks required for implementing arbitrary unitary transformations scales only linearly with the number of manipulated modes, making this approach potentially scalable\cite{morizur2010programmable,labroille2014efficient,fontaine2017design,LopezPastor21}. Recently, MPLCs have been used for performing unitary gates on single photons in high-dimensions\cite{brandt2020high,li2020programmable} and for studying two-photon interference of separable orbital angular momentum (OAM) states\cite{hiekkamaki2021high}, showing their compatibility with quantum states of light. It is thus appealing to consider MPLC as a platform for processing high-dimensional entanglement.

In this work, we demonstrate the applicability of MPLC as a versatile and scalable platform for quantum information processing of entangled photons in high dimensions. We study four key tasks for quantum information processing: high-dimensional entanglement certification, tailored two-photon interference, arbitrary random transformations and spatial mode conversions for entanglement distribution. All tasks are performed by simply changing the reconfigurable phase patterns employed by the MPLC, without any changes in the hardware. To certify two- and three-dimensional entanglement and observe tailored two-photon interference, we program the MPLC to switch between mutually unbiased bases (MUBs)\cite{bavaresco2018measurements}. To demonstrate the universality of MPLC processing we experimentally realize 400 Haar random unitary transformations and verify both the fidelity and the statistics of the random output states\cite{carolan2015universal,flamini2018photonic}. Finally, we demonstrate mode conversion between the original path-encoded state and a linearly polarized (LP) mode basis that is suitable for entanglement distribution via a few-mode fiber link\cite{cirac1997quantum,steinlechner2017distribution,hu2020efficient,cao2020distribution}. These results pave the way toward both fundamental research and technological applications of high-dimensional entanglement utilizing MPLC as a universal entanglement processor.

\section{Results}

\begin{figure}[H]
\centering
\includegraphics[width=0.9\textwidth]{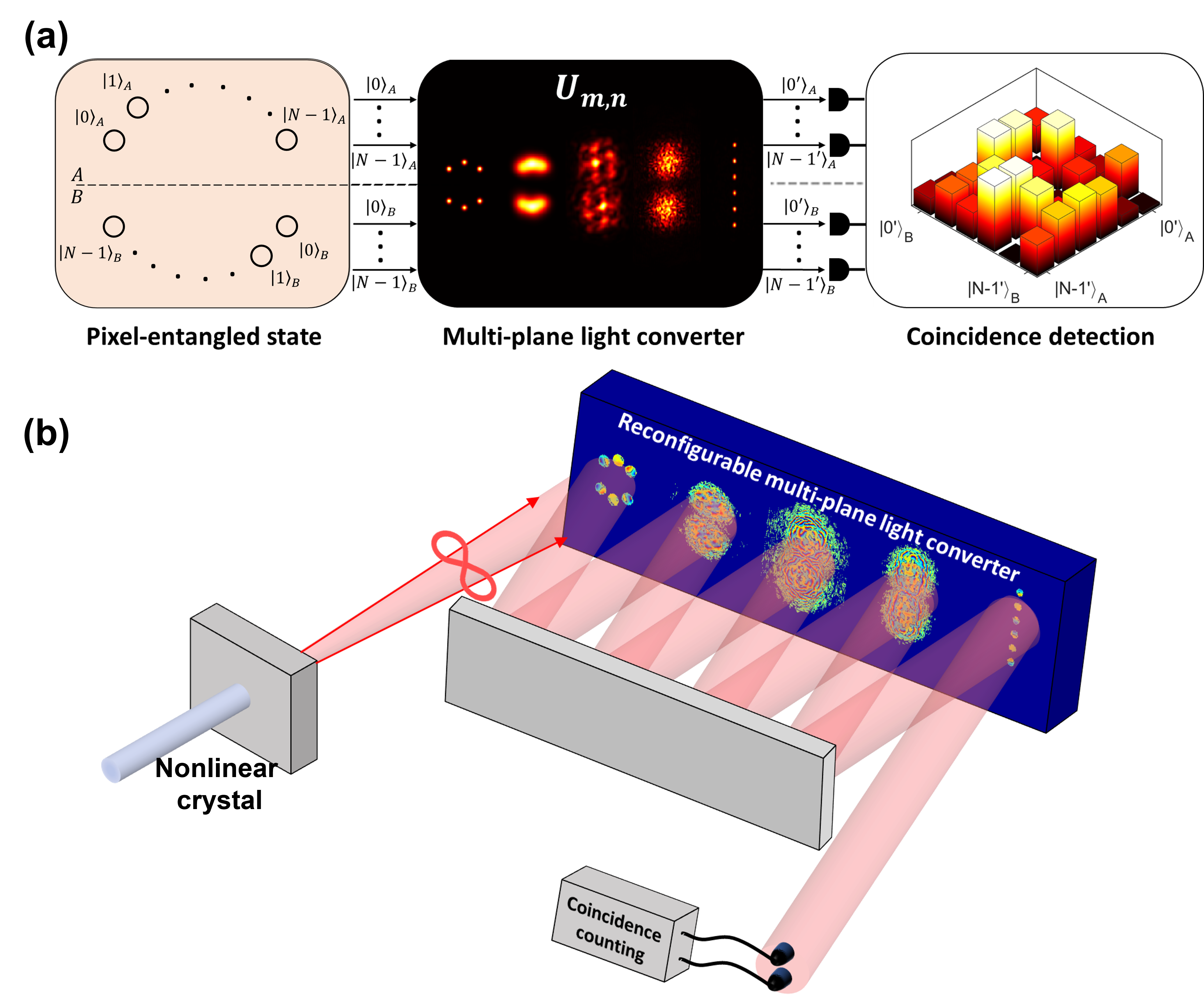}
  \caption{\label{fig:1} \textbf{Processing high-dimensional entanglement.} (a) A pixel-entangled two-photon state undergoes a general unitary transformation implemented using a multi-plane light converter (MPLC). The output quantum correlations are measured at the last plane of the MPLC. (b) An illustration of the experimental implementation. Pixel-entangled photons are generated via spontaneous parametric down conversion. The MPLC is placed at the far-field of the nonlinear crystal and consists of five planes, implemented by successive reflection between a phase-only spatial light modulator and a mirror. The correlations between the photons at different transverse positions are measured at the output using two single photon detectors placed on motorized stages that scan their transverse position.}
\end{figure}

The concept of processing high-dimensional entanglement using an MPLC is depicted in fig. \ref{fig:1}a. The two-photon entangled state is encoded using $2N$ spatial modes in the so-called pixel basis\cite{o2005pixel,valencia2020high}. The $N$ upper (lower) modes are labeled as $\ket{0}_A,\cdot \cdot \cdot,\ket{N-1}_A$ ($\ket{0}_B,\cdot \cdot \cdot,\ket{N-1}_B$). The desired unitary operation is programmed onto the MPLC which transforms the input state accordingly, and the correlations between different output modes are then measured using single photon detectors. 

Experimentally, we generate the spatially entangled input state via spontaneous parametric down conversion (SPDC) by pumping a nonlinear crystal with a pump laser beam (fig. \ref{fig:1}b). The MPLC is located at the far-field of the crystal, thus, due to transverse momentum correlations in SPDC, the input quantum state is spatially entangled and is given approximately by $\ket{\psi} = \frac{1}{\sqrt{N}} (\ket{0}_A\ket{0}_B+ \cdot \cdot \cdot +\ket{N-1}_A\ket{N-1}_B)$. 

We note that the above quantum state is presented in the Schmidt mode basis, which is highly convenient for entanglement certification\cite{bavaresco2018measurements}. For pixel entangled states generated via SPDC, the pixel and Schmidt bases are equivalent under two main conditions, which are fulfilled in our experiment: the defined pixels are separated by more than the width of the pump beam, leading to minimal coincidence crosstalk, and are also within the range of the phase matching condition of the SPDC process\cite{srivastav2021characterising}.
We experimentally implement a five-plane MPLC by successive reflections of the photons between a phase-only spatial light modulator (SLM) and a mirror. The algorithm for finding the required phase masks for a desired unitary transformation is based on the wavefront-matching algorithm\cite{fontaine2019laguerre} (see Appendix B).

\subsection{Entanglement certification}
We begin by using the MPLC to certify high-dimensional entanglement\cite{bavaresco2018measurements,friis2019entanglement}. To certify entanglement in two or more dimensions, one must measure the quantum correlations between the photons in at least two MUBs\cite{bavaresco2018measurements,wootters1989optimal}. In our case, we utilize the reconfigurability of the MPLC to switch between measurements in the 'standard' $\ket{0},\cdot \cdot \cdot,\ket{N-1}$ pixel basis, and in the discrete Fourier transform (DFT) MUB defined by $\ket{j'}_{A/B}=\frac{1}{\sqrt{N}} \sum_{m=0}^{N-1}\omega^{jm}\ket{m}_{A/B}$, where $N$ is the dimension of the Hilbert space of each photon and $\omega=exp(2\pi i/N)$. It has recently been proved that this set of measurements is sufficient for obtaining a lower bound on the fidelity $F$ between the inspected quantum state and a maximally entangled target state\cite{bavaresco2018measurements}. Using this bound, m-dimensional entanglement within an N-dimensional state is certified for $F>(m-1)/N$\cite{bavaresco2018measurements}.

\begin{figure}[H]
\centering
\includegraphics[width=0.8\textwidth]{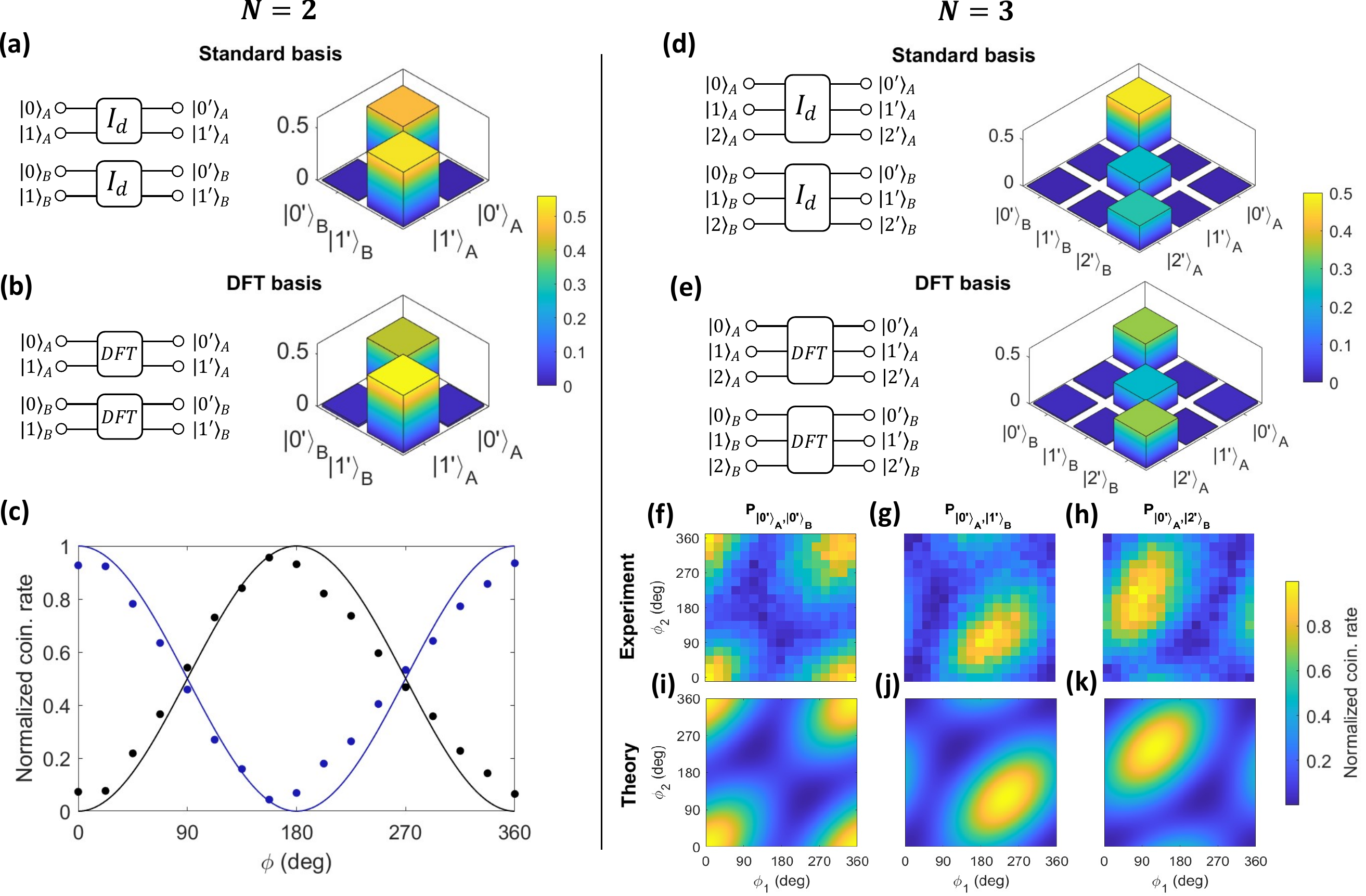}
  \caption{\label{fig:2}  \textbf{Entanglement certification.} A maximally entangled Bell state is encoded by considering the spatially entangled photons in two input spatial modes per photon. To test the fidelity of our state with respect to the maximally entangled Bell state, measurements in the standard basis (a) and in the DFT mutually unbiased basis (MUB) (b) are performed by programming the desired transformations using the MPLC. A fidelity $F \geq 95\pm 1\%$ is obtained, certifying entanglement. A relative phase $\phi$ in the input state is scanned by adding the desired phase to one of the modes at the input plane of the MPLC, yielding the expected sinusoidal change in the coincidence rate (c). The small observed shift between the theoretical prediction (not fitted) and the measured data, might result from a constant phase shift between the input modes, which is likely caused by slightly imperfect alignment or phase correction of the SLM. The error bars are smaller than the size of the data points. Next, we consider three input spatial modes per photon and measure the fidelity of our state compared with the three-dimensional entangled state. Programming the MPLC to switch between the standard (d) and the DFT mutually unbiased basis (e), a fidelity $F \geq 90 \pm 2 \%$ is obtained, certifying genuine three-dimensional entanglement. Similarly to the two-dimensional case, the input plane of the MPLC can be used to scan the two relative phases between the terms. The experimental results in (f-h) and the theoretical predictions in (i-k) for the correlations between the different output modes are in good agreement, exhibiting quantum interference with high visibility.}
\end{figure}

In two dimensions, we measure the fidelity of the state compared with the maximally entangled Bell state $\ket{\Phi^+}=\frac{1}{\sqrt{2}} (\ket{0}_A\ket{0}_B+\ket{1}_A\ket{1}_B)$. We first program the MPLC to perform the identity operation, which maps each input spot to a single output spot. Strong correlations between the photons are obtained in this basis (fig. \ref{fig:2}a). To obtain a lower bound for the fidelity, we program the MPLC to switch to the DFT basis defined above, where strong correlations are observed as well (fig. \ref{fig:2}b). Using these two results, we certify the entanglement by obtaining a lower bound for the fidelity  $F\geq 95\pm 1\%$, significantly above the upper bound for separable states of $50\%$\cite{bavaresco2018measurements}. To further quantify the performance and coherence of the transformation, we look at the visibility of the quantum interference in the DFT basis, by applying a phase $\phi$ to one of the spots at the input plane. The quantum correlations at the output change sinusoidally with respect to this relative phase, in agreement with the theoretical prediction and with a high visibility of $94.3 \pm 0.4\%$ (fig. \ref{fig:2}c).

In three dimensions, we perform similar measurements to obtain a lower bound for the fidelity of our state with respect to $\ket{\psi}=\frac{1}{\sqrt{3}} (\ket{0}_A\ket{0}_B+\ket{1}_A\ket{1}_B+\ket{2}_A\ket{2}_B)$. By measuring the correlations in the standard and DFT bases, we certify genuine three-dimensional entanglement by obtaining $F \geq 90 \pm 2 \%$ (fig. \ref{fig:2}d,e), which is above the upper bound of $67\%$ obtainable with two-dimensional entanglement\cite{bavaresco2018measurements,friis2019entanglement}. In three dimensions, two relative phases between the different terms in the quantum state, $\phi_1,\phi_2$, can affect the quantum interference between the photons in the DFT basis. By varying the phases of two input spots and measuring the coincidence rate between different output modes we observe quantum interference with an average visibility $93.5 \pm 0.5\%$ (fig. \ref{fig:2}f-h), and in good agreement with the theoretical prediction (fig. \ref{fig:2}i-k).

\subsection{Haar random transformations}
\begin{figure}[H]
\centering
\includegraphics[width=0.7\textwidth]{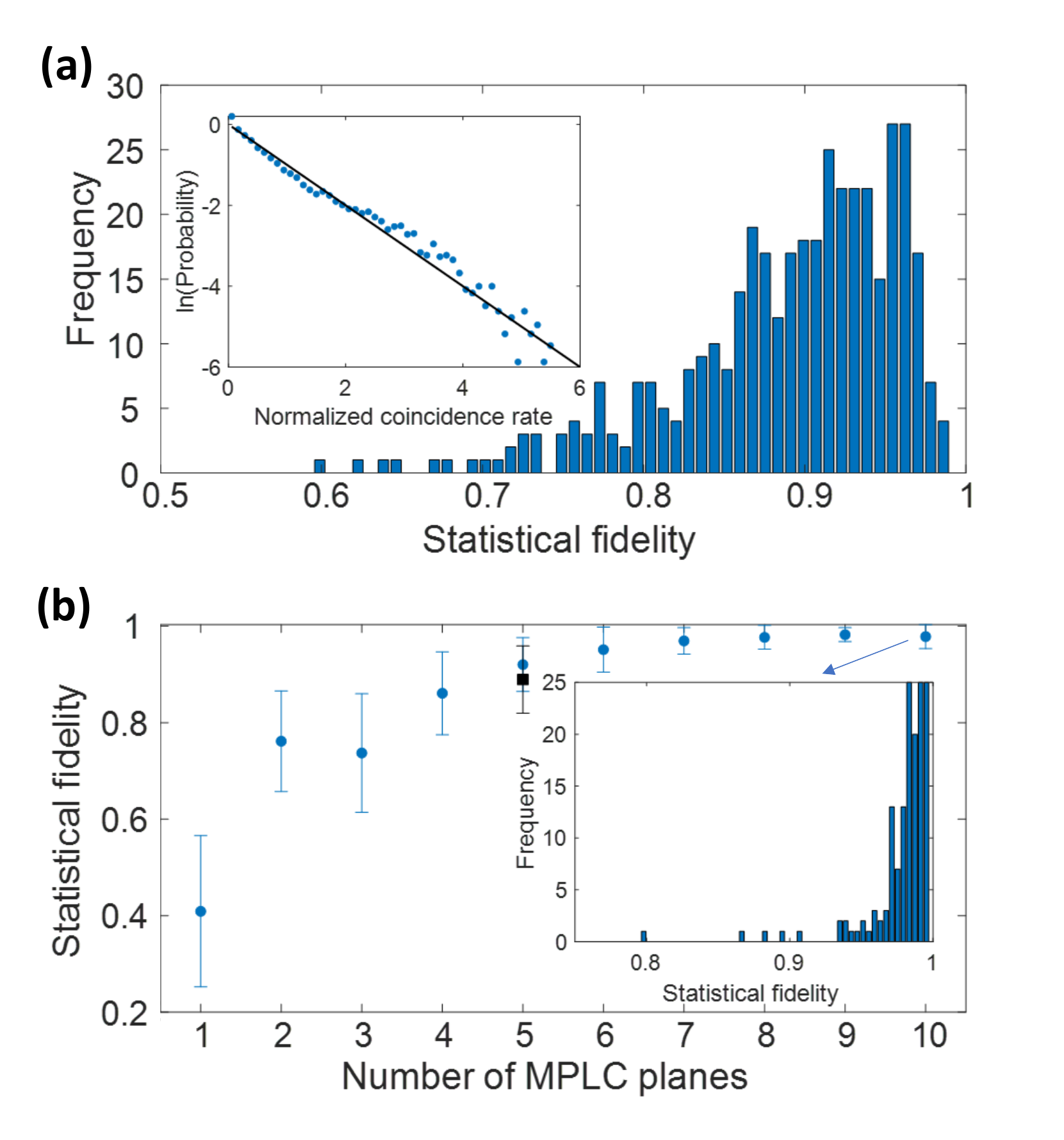}
  \caption{\label{fig:3}  \textbf{Programming arbitrary random transformations.} We use the MPLC to experimentally implement 400 random unitary transformations on four input modes, sampled according to the Haar measure. A histogram of the obtained statistical fidelities is presented in (a), showing an average statistical fidelity $F_s=88.9\pm 0.3\%$. In the inset, the probability distribution of the coincidence rates between the different modes normalized by their mean is presented (blue points) together with the theoretical prediction according to the Porter-Thomas distribution (black line). In (b), a simulation of the average statistical fidelity for four-mode transformations as a function of the number of planes in the MPLC is presented (blue points). Our experimental result with five planes is marked with a black square. The histogram of the simulated statistical fidelities for an MPLC with ten planes is presented in the inset, exhibiting a statistical fidelity of $98\pm 2\%$.}
\end{figure}

Next, we showcase the universality and reconfigurability of the MPLC platform by programming Haar random unitary matrices, which are of great interest for applications such as boson sampling\cite{zhong2020quantum,carolan2015universal} and quantum cryptography\cite{hayden2004randomizing}. We experimentally implement 400 random four-dimensional unitary transformations on four spatial modes (two modes per photon), sampled according to the Haar measure\cite{zyczkowski1994random}. For each random unitary, the coincidence rates between all pairs of output modes are measured. The probability distribution of the coincidence rates for different unitary transformations is presented in the inset of fig. \ref{fig:3}a. For random transformations, according to the Porter-Thomas distribution, a negative exponential distribution is theoretically expected (black line)\cite{boixo2018characterizing,porter1956fluctuations}, with good agreement with the experimental results (blue dots).
As a measure of the resemblance between the experimentally measured correlations between each pair of modes $P_i^{exp}$ and the theoretical predictions $P_i^{th}$ for each unitary transformation, we utilize the commonly used statistical fidelity defined as $F_s=\sum_i\sqrt{P_i^{exp}P_i^{th}}$\cite{carolan2015universal}. A histogram of the 400 statistical fidelity values is presented in fig. \ref{fig:3}a. The average statistical fidelity obtained in our experiment is $F_s = 88.9\pm 0.3\%$. The fidelity can be improved even further by increasing the number of planes in the MPLC. To show this, we simulate the performance of the MPLC for random transformations as a function of the number of planes (fig. \ref{fig:3}b). Statistical fidelities of up to $98\pm 2\%$ can be obtained by as few as ten planes. The high statistical fidelities that we obtain between the MPLC generated transformations and the target Haar transformations, manifest the capability of MPLC to generate arbitrary unitary transformations, as by definition Haar matrices uniformly sample the space of all unitary transformations\cite{taballione2021universal}.

\subsection{Mode conversion}
\begin{figure}[H]
\centering
\includegraphics[width=0.5\textwidth]{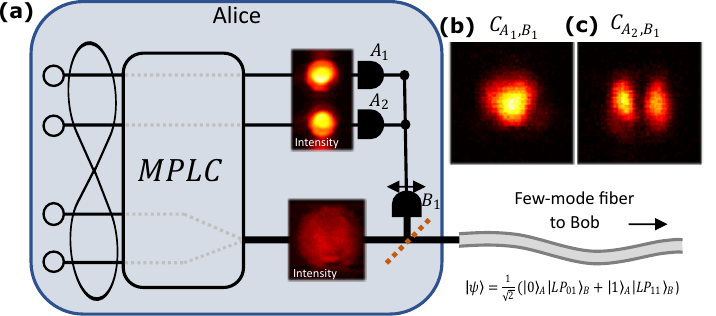}
  \caption{\label{fig:4}   \textbf{Mode conversion for entanglement distribution.} (a) An illustration of entanglement distribution utilizing the MPLC as a mode converter. The MPLC is programmed such that one of the photons is left untouched and stays with Alice, while the other is transformed from the pixel basis to a fiber mode basis and can be transmitted to Bob via a few-mode fiber link. In the experiment, we transform Bob's photon to a $LP_{01}, LP_{11}$ basis and measure its spatial distribution conditioned on the detection of a photon at Alice's top (b) or bottom (c) spots.}
\end{figure}

Finally, we demonstrate the relevance of MPLC for entanglement distribution\cite{cirac1997quantum,steinlechner2017distribution,hu2020efficient,cao2020distribution}. The encoding of entangled photons in modes compatible with the quantum link used for their distribution is important for reducing loss and crosstalk, making mode converters central for such applications. Here, we consider the case where Alice generates a pair of photons entangled in the pixel basis, and wishes to distribute one photon to Bob via a few-mode fiber link. Alice can thus use her MPLC to convert Bob's photon to the LP mode basis supported by the weakly guided optical fiber, while keeping her photon in the pixel basis for ease of detection (fig. \ref{fig:4}a). In the experiment, we consider the initial state $\ket{\Phi^+}=\frac{1}{\sqrt{2}} (\ket{0}_A\ket{0}_B+\ket{1}_A\ket{1}_B)$ generated by Alice, and program the MPLC to convert Bob's photons such that the state $\ket{\psi}=\frac{1}{\sqrt{2}} (\ket{0}_A\ket{LP_{01}}_B+\ket{1}_A\ket{LP_{11}}_B)$ is obtained (see Appendix B). By keeping Alice's detector at a fixed position and scanning the position of Bob's detector at the output of the MPLC, we observe a correlation between the LP mode detected and the position of Alice's stationary detector (fig. \ref{fig:4}b,c). 

\section{Conclusion}
In conclusion, we have discussed and demonstrated the applicability of MPLC as a universal and scalable platform for processing high-dimensional entanglement. We have certified high-dimensional entanglement with high fidelity by programming the MPLC to switch between different mutually unbiased bases\cite{bavaresco2018measurements,brandt2020high}. By scanning the phases of the spots at the input plane and measuring in the DFT basis, we have observed high-visibility quantum interference, which is fundamental to many quantum information processing tasks. Furthermore, to demonstrate the universality of MPLC, we have experimentally programmed 400 Haar random unitary transformations on four modes using MPLC with five planes, observing high statistical fidelity\cite{carolan2015universal}. Besides demonstrating the reconfigurability and universality of the MPLC, Haar random unitaries are important for boson sampling, quantum key distribution and other quantum information processing tasks\cite{zhong2020quantum,carolan2015universal,flamini2018photonic,hayden2004randomizing}. Finally, we have considered the use of MPLC as a mode converter for entanglement distribution, demonstrating correlations between the position of one photon and the measured LP mode of the other. 

We note that the use of SLMs and phase plates for performing quantum information processing tasks became a common practice in recent years, especially for high-dimensional states\cite{erhard2020advances}. For example, two-photon quantum interference and single photon high-dimensional gates were recently realized using MPLC in the OAM degree of freedom\cite{brandt2020high,hiekkamaki2021high}, OAM- and pixel-entanglement certification were demonstrated with the aid of SLMs\cite{bavaresco2018measurements,valencia2020high}, and mode converters in both the classical and quantum domains have been realized using static phase plates\cite{cui2017distribution} and SLMs\cite{fontaine2017design}. Our work contributes to this ongoing effort by demonstrating the manipulation of high dimensional entanglement using MPLC and by performing a variety of different tasks using a single, reconfigurable platform, rather than a specific setup for each given task. 

Being a reconfigurable, scalable and universal approach for processing entangled photons in high dimensions that is also compatible with structure light, MPLC can serve as a leading platform for both fundamental and applied quantum information science. We also anticipate that the wide bandwidth of MPLC\cite{fontaine2020ultrabroadband} together with the ability to control all degrees of freedom of light\cite{mounaix2020time}, will open possibilities in controlling hyper entangled states\cite{deng2017quantum}. Finally, as recent results in the classical domain demonstrated the use of MPLC for sorting over a thousand spatial modes with only 14 phase masks\cite{fontaine2021hermite}, we believe MPLC could be used for processing entangled photons encoded in record-high dimensions in the near future.

\section*{Acknowledgments}
This work is supported by the Zuckerman STEM Leadership Program, the Israel Science Foundation (grant No. 1268/16), the United States-Israel Binational Science Foundation (BSF) (Grant No. 2017694) and by the ISF-NRF Singapore joint research program (grant No. 3538/20).

\section*{Appendix A: Detailed experimental setup}
\begin{figure}[H]
\centering
\includegraphics[width=0.8\textwidth]{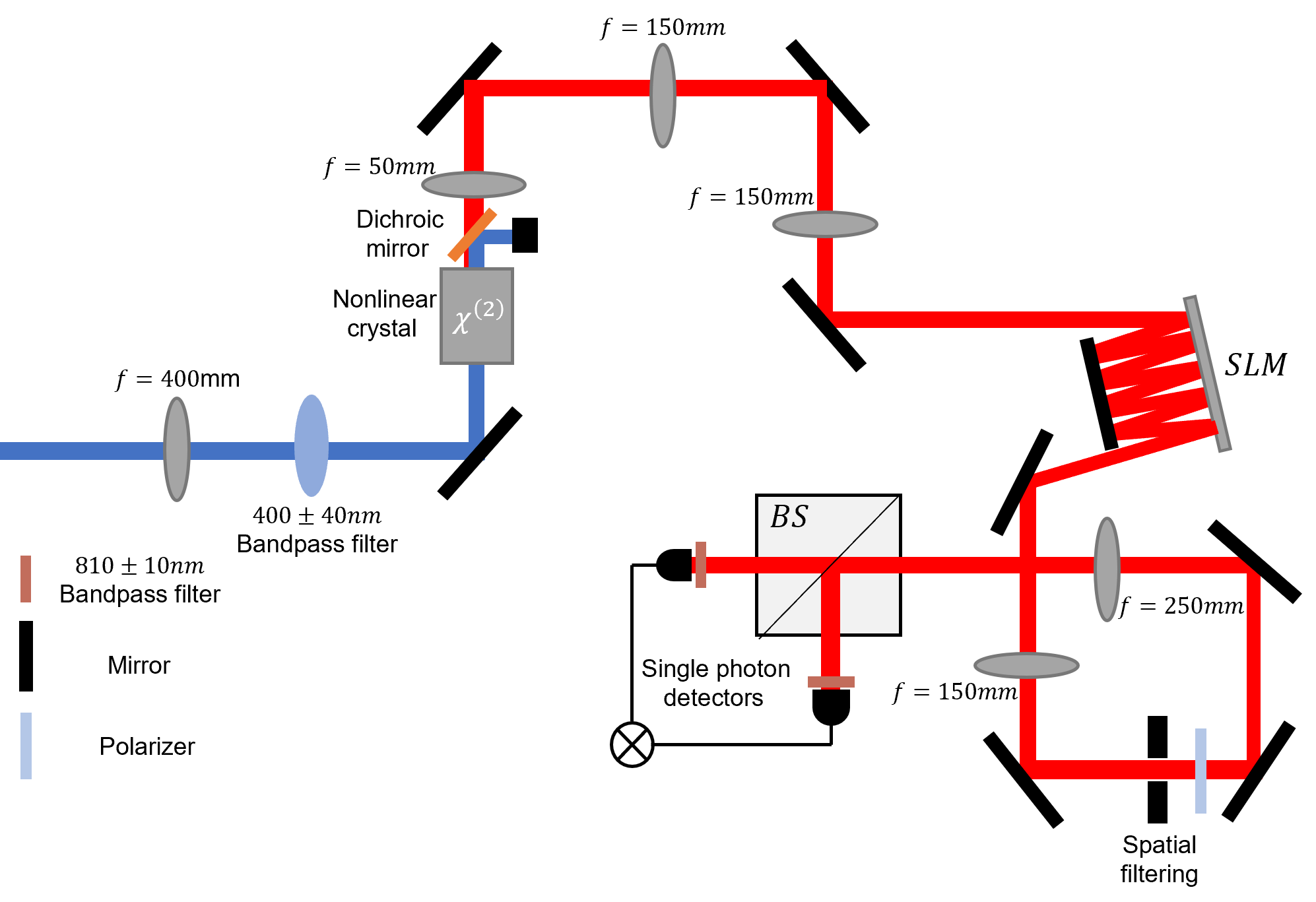}
  \caption{\label{fig:s1}  Detailed experimental setup.}
\end{figure}

A detailed illustration of the experimental setup is given in fig. \ref{fig:s1}. A $250mW, 405nm$ continues wave laser is used as a pump beam for the spontaneous parametric down conversion (SPDC) process. The pump beam is first filtered by two $40nm$ filters centered at $400nm$ to exclude any residual light at undesired wavelengths, and then focused by a $400 mm$ lens onto an $8mm$ thick type-I BBO crystal, exhibiting a waist of $55 \mu m$ at the crystal plane. After the crystal, a dichroic mirror is used to separate the pump beam from the entangled photons. A 2-f system with a focal length of $50 mm$ followed by a 4-f system with a focal length of $150 mm$ are used so that the far-field distribution of the photons is imaged onto the first plane of the MPLC. 

The MPLC consists of five planes, implemented by reflecting the entangled photons between a liquid crystal SLM (pixel size $12.5 \mu m$) and a dielectric mirror. Free space propagation of $76 mm$ separates the different planes in the MPLC. A vertical linear phase is implemented in all areas of the SLM not used for the transformation, to allow efficient blocking of residual light by a spatial filter located after the MPLC, at the Fourier plane of the last phase mask. As the SLM modulates only horizontally polarized light, we place a polarizer at the output of the MPLC to eliminate any unmodulated light. Lenses with focal lengths $150 mm$ and $250 mm$ are used to image the final plane of the MPLC onto two $100 \mu m$ multimode fibers connected to single photon detectors. A $50/50$ beamsplitter is used to probabilistically split the light into the two detectors, which are filtered with $10 nm$ filters centered around $810 nm$. Coincidence counts are recorded using a dedicated FPGA, with a coincidence window of $4ns$. Accidental counts are subtracted in all coincidence measurements.

\section*{Appendix B: Methods}
\textbf{Wavefront matching}. The required phase masks for every transformation presented in this work were calculated by the wavefront matching optimization method. Given a set of input and output modes defining the desired unitary transformation, an optimization process with between 20 and 50 iterations has been performed. In every iteration, each phase mask is adjusted by propagating all input modes and back-propagating the output modes to the same relevant plane, and taking their phase differences into account. Thanks to reciprocity, for a perfect transformation, the output and input modes must have the same phase profile at each plane. The algorithm thus adjust the phase at each plane so that the average error between all input and output modes is minimal (see\cite{fontaine2019laguerre} for more details and an open-access Matlab code). To avoid scattering losses and obtain smoother phase masks, we have limited the optimization algorithm to utilize angles up to only a quarter from the maximal diffraction angle determined by the pixel size of the SLM. In addition, utilizing the fact that our modes are spatially separated at the input plane, an additional phase was added to each input mode to ensure their global phases at the output are correct.

\noindent \textbf{LP modes}. The field distributions of the output $LP_{01}$ and $LP_{11}$ target modes in fig. 4 were calculated for a step index fiber with a diameter of $50 \mu m $ and numerical aperture of $0.2$, at the wavelength of $808nm$ \cite{agrawal2012fiber}. The size of the actual target output modes was scaled up by a factor of ten to allow measuring them with good spatial resolution.

\section*{Appendix C: MPLC efficiency}
We define the efficiency of the MPLC as the ratio between the total intensities measured at the output and input modes of the MPLC. To estimate the efficiency of the MPLC, we consider 50 Haar random transformations on four spots (as in fig. 3), and compare for each transformation the total intensity at the relevant spots before and after the MPLC. We observe an average efficiency of $19 \pm 5\%$, which is comparable with recent demonstrations with classical light\cite{fontaine2019laguerre}. The efficiency of the MPLC in our experiment is reduced mainly due to loss ($17 \%$), the accuracy of the transformations which according to simulations reduce the efficiency by $31 \pm 8 \%$, and the imperfect diffraction efficiency of the SLM. We note that in principal, MPLCs with much higher efficiencies can be achieved, using for example deformable mirrors instead of liquid-crystal SLMs.

\bibliography{sample}

\bibliographystyle{naturemag}

\end{document}